# Ultrafast image retrieval from a holographic memory disc for high-speed operation of a shift, scale, and rotation invariant target recognition system

**Julian Gamboa**
*Department of ECE, Northwestern University, Evanston, IL, 60208, USA*
**Xi Shen**
*Department of ECE, Northwestern University, Evanston, IL, 60208, USA*
**Tabassom Hamidfar**
*Department of ECE, Northwestern University, Evanston, IL, 60208, USA*
**Selim M. Shahriar**
*Department of ECE, Northwestern University, Evanston, IL, 60208, USA*
*Department of Physics and Astronomy, Northwestern University, Evanston, IL, 60208, USA*

**Abstract:** The hybrid opto-electronic correlator (HOC) architecture has been shown to be able to detect matches in a shift, scale, and rotation invariant (SSRI) manner by incorporating a polar Mellin transform (PMT) pre-processing step. Here we demonstrate the design and use of a thick holographic memory disc (HMD) for high-speed SSRI correlation employing an HOC. The HMD was written to have 1,320 stored images, including both unprocessed images and their PMTs. We further propose and demonstrate a novel approach whereby the HOC inputs are spatially shifted to produce correlation signals without requiring stabilization of optical phases, yielding results that are in good agreement with the theory. Use of this approach vastly simplifies the design and operation of the HOC, while improving its stability significantly. Finally, a real-time opto-electronic PMT pre-processor utilizing an FPGA is proposed and prototyped, allowing for the automatic conversion of images into their PMTs without additional processing delay.

## 1. INTRODUCTION

High-speed target recognition is of significant importance in many arenas. In space situational awareness (SSA), for example, it is essential to be able to identify a target quickly so as to not cause delays in making a crucial decision. From applications as diverse as military and social, the need to identify shapes, images, and signatures has led to the widespread search for techniques that may save processing time and resources. To this end, the field of optics has provided many incarnations of correlators thanks to the inherent properties of convex lenses which allow them to perform the 2D Fourier transform (FT) of an image that is passed through them. In particular, if a collimated image beam is placed at the focal plane of a lens, then the FT of the image will be generated at the opposite focal plane, thus performing the FT in the time it takes light to propagate a short distance [1]. This property, together with the well-known relationship between products of FTs and correlation/convolution, make optical techniques eminently suited for high-speed target recognition. The Vander-Lugt [1] and joint-transform [2] correlators are perhaps the two best known examples of how this property may be exploited for all-optical correlation. However, these techniques are limited in their capacity for real-time operation, as they require a pre-recording step or sensitive nonlinear materials, respectively [3].

In recent years we have shown an alternative hybrid opto-electronic correlator (HOC) architecture that combines high-speed optical FTs with high-speed electronic processing to achieve target recognition [4,5]. While all FT correlators are inherently shift invariant, further work is required in order to achieve scale and rotation invariance [6,7]. To this end we have further demonstrated the successful incorporation of the polar Mellin transform (PMT) in the HOC architecture to achieve shift, scale, and rotation invariant (SSRI) image correlation [8,9].

The HOC architecture contains two image arms: one for a reference (Ref) and another for a query (Que) image. In a simplified version of the HOC, both arms utilize a spatial light modulator (SLM) to project the desired image into the optical domain. While this is convenient for testing and research, it results in a severely limited operational speed. A more complete version of the HOC replaces the reference SLM with a holographic memory disc (HMD) that contains a large database of images which can be read out at the speed of light. Such an all-optical storage device is inherently superior to electronic alternatives for this particular application in that it does not require any additional processing time to select the image. This leads to a unique zero-latency property that sets it apart from other modern storage media. Phenanthrenequinone (PQ) doped poly(methyl methacrylate) (PMMA) is a well-known write-once read-many (WORM) holographic substrate that has been at the center of thick holography research for over two decades [10–12]. The properties of this material have been studied in detail in regard to storage capacity [12–18], storage modes [19–24], and manufacturing techniques [10,11,25–28].

The rest of the paper is structured as follows. Section 2 details the properties and use of PQ:PMMA HMDs. An overview of SSRI target recognition is provided in section 3. A novel FPGA-based opto-electronic method for generating the PMT of an image is detailed in section 4. Experimental results of HOC correlations using a high-capacity HMD are presented in section 5. Finally, conclusions and our outlook are given in section 6.

## 2. PQ:PMMA HMDs AS ULTRA-FAST ALL-OPTICAL IMAGE DATABASES

PQ:PMMA is a WORM holographic substrate that has been well studied for a variety of applications ranging from data storage to wavelength filtering. The WORM aspect of this material is especially useful for large databases, as it ensures minimal degradation of the stored data from one read cycle to another [11,28]. The medium is composed of a polymer matrix doped with a photosensitive dye. When the dye is exposed to light, it is elevated to a higher energy state and consequently attaches itself to PMMA and MMA, forming oligomers with a volumetric density distinct from the host matrix [11,28]. This property can be exploited by selectively exposing regions in order to generate a non-uniformity of the refractive index of the material. In this way, holographic gratings are formed when two coherent beams of light are interfered within the material. Due to the aforementioned photosensitivity of the PQ, the pattern will be directly imprinted onto the substrate, resulting in volumetric spatial modulation of the refractive index ($\Delta n$). This material is unique among holographic substrates in that the polymerization occurs independently of the exposure, and thus experiences minimal shrinkage after writing. This is accomplished through the use of azobisisobutyronitrile (AIBN) to pre-polymerize the PMMA during fabrication. However, this has the additional effect of decreasing the maximum $\Delta n$ as compared to other holographic polymers. To compensate, PQ:PMMA holographic optical elements (HOE) must be fabricated much thicker, typically in the millimeter range, in order to achieve useful diffraction efficiencies (DE), pushing the gratings into the Bragg regime. In the case of plane-wave writing beams, the resulting grating will be perfectly sinusoidal, and is represented by a grating vector which can be expressed as $\boldsymbol{K}_g = \boldsymbol{K}_1 - \boldsymbol{K}_2$, where $\boldsymbol{K}_g$ is the grating wave vector and $\boldsymbol{K}_{1,2}$ are the wave vectors of the two writing beams [29]. Because this condition is highly specific, each combination of writing wave vectors would yield a distinct grating and so can be exploited to store large amounts of images at a single location, as is shown in section 2.1 below. The capacity of an HMD will of course depend on many parameters but can generally be estimated through the value of $\Delta n$, the sample thickness, and a parameter known as the M/#. The latter is a representation of the ratio of the maximum $\Delta n$ and the value of $\Delta n$ that yields 100% DE in an HOE [16,30]. For this material, a value of M/# as high as 9.4 has been reported previously [16].

### 2.1 STORING A LARGE IMAGE DATABASE ON A PQ:PMMA HMD

Holograms written in PQ:PMMA are typically a few millimeters in thickness, and so fall into the Bragg regime [20]. The Bragg condition can be extended to non-planar waves by defining each beam to be a superposition of an infinite set of perfect plane waves through Fourier theory. In this way, we can replace one of the plane-wave writing beams with an image beam to form a complex grating with a unique Bragg condition. Furthermore, because of the highly selective nature of Bragg gratings, it is possible to write overlapping holograms at a single location so long as they possess distinct Bragg conditions [19–22,29]. To do this, we may vary the writing wave vectors either through their angles or wavelengths. The wavelength approach is limited due to the spectral range of the photosensitivity of the material, which is limited to wavelengths up to ~532 nm, while also requiring extremely complex alignment setups. For this reason, angle multiplexing is typically preferred, where the input angles of the writing beams are changed from one grating to another. Thus, HOEs can be used as HMDs by designing writing and reading systems that take into account these properties. To do this, each hologram will have to be spaced far enough from neighboring holograms so as to avoid crosstalk. From Kogelnik's coupled wave theory [29], the angular bandwidth of a thick unslanted phase hologram, that is to say the angular deviation from the Bragg condition that yields half of the maximum DE, can be expressed as $\Delta \theta_{1/2} \approx \Lambda_g / 2d$, where $\Lambda_g$ is the wavelength of the grating and $d$ is the thickness of the substrate.

We may choose the angular spacing, $\Delta \theta$, to be much larger than $\Delta \theta_{1/2}$, which would yield high quality holograms, albeit at a suboptimal capacity. Alternatively, we can analyze the behavior of the DE in holograms to find an optimum grating spacing. The DE for lossless phase gratings can be found to behave as a sinc function through Kogelnik's coupled wave theory [29]. Thus, it is optimal to position gratings such that the first zero of one hologram overlaps the peak of the adjacent hologram in a manner similar to that of orthogonal frequency division multiplexing in telecommunications [31]. However, the non-planar aspect of real laser beams will result in a larger angular and spectral bandwidth than that of plane-wave gratings. As such, this approach will only yield an estimate of the optimal positions. The initial estimate can be improved by experimentally measuring the angular DE profile of image holograms to find the true value of $\Delta\theta_{1/2}$.

When writing image holograms, there have traditionally been two techniques depending on the precise application: either the image itself [8], or its FT [22] can be written to the HMD. The former requires more precise alignment of the optical writing system and will occupy more physical space on the disc but will also reproduce the image directly when read. The latter requires a larger $\Delta\theta$ to compensate for the lack of collimation of the beam but uses less space at each writing location and is less susceptible to imperfections in the substrate. For the particular case of the HOC, it is convenient to perform direct-image writing in order to simplify image retrieval while correlating signals.

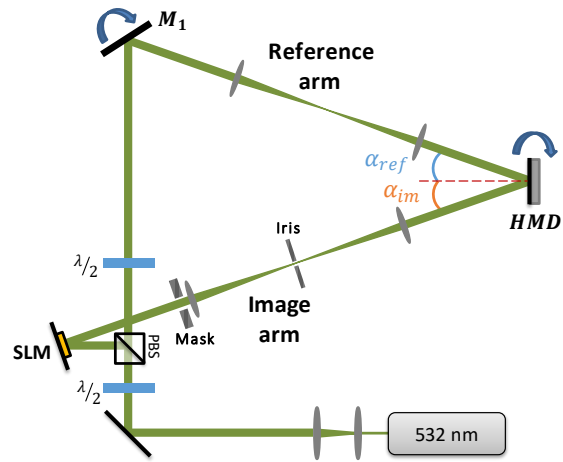

Fig. 1. HMD writing setup. A 532 nm beam is expanded and split into a reference arm and an image arm. The reference arm is composed of a motorized rotating mirror ($M_1$) and a 4-f system which together control $\alpha_{ref}$. The image arm is composed of an SLM, a mask, a 4-f system, and an iris. The HMD is placed at the output of both of the 4-f systems, and is itself mounted on a motorized rotation stage which allows it to change $\alpha_{im}$ and $\alpha_{ref}$. A PBS and two HWPs control the relative power of both arms.

Fig. 1 shows a diagram of the constructed image writing setup which utilizes a digital micromirror device (DMD) SLM to project a digital image onto the spatial intensity profile of a collimated 25mm diameter beam. Here, a 532nm laser beam is expanded to 25mm and passed through a half wave plate (HWP) which is used to control the relative intensities of the two arms. The beam passes through a polarized beam splitter (PBS) into the image arm and the reference arm, where a second HWP corrects the polarization of one of the two so as to allow them to interfere. The reference arm consists of a motorized mirror, $M_1$, placed at the input plane of a 4-f system, with the HMD placed at the output plane. This configuration enables us to control the input angle of the reference arm without changing the position of the HMD, as explained in [20]. The image arm replaces the rotating mirror with a DMD SLM and utilizes an identical 4f system to limit the adverse effects of free-space diffraction on the image quality. It is important to note that this type of SLM is composed of millions of microscopic mirrors that shift in order to redirect light either towards or away from the target. These mirrors are themselves separated a small distance from the nearest neighbors to avoid mechanical collisions, and so contain a grid of edges from which light diffracts. This effect generates many higher order diffractions that get in the way of the holographic recording process. For this reason, a mask is placed in front of the arm's 4f system, and an iris is incorporated at the Fourier plane. The iris must be tuned to allow as much of the SLM's reflection ($0^{th}$ order) to pass through, while blocking all of the higher order diffractions. The HMD itself is also placed upon a motorized rotation stage so as to be able to control the input angle of the image beam. We denote the angle of $M_1$'s motorized rotation stage by $\angle M_1$, and the angle of the HMD's motorized rotation stage by $\angle HMD$. $\angle M_1$ is set to 0° at the position where the reference beam passes through the center of the lenses in its respective 4-f system. $\angle HMD$ is set to 0° at the position where, with $\angle M_1 = 0°$, $\alpha_{ref}$ and $\alpha_{im}$ have equal values. The writing angles $\alpha_{ref}$ and $\alpha_{im}$ are defined as the angles between the normal of the HMD's front surface and the reference or image beam, respectively, and are always taken to be positive.

Image multiplexing is accomplished by first loading an image onto the SLM, exposing the HMD for a set amount of time, and subsequently changing one or more of the involved angles prior to exposing the next image. In the case of the HOC, there are two principal options for angle control, depending on the desired reading mode. In the first mode the rotating mirror's angle is unchanged, and the entirety of the HMD is rotated. In this mode, when the HMD is rotated clockwise by $\Delta\angle HMD$, the writing angles relative to the hologram's normal are changed from their initial positions by $\Delta\alpha_{ref} = -\Delta\angle HMD$ and $\Delta\alpha_{im} = \Delta\angle HMD$, respectively, thus generating radically different gratings for each value of $\Delta\angle HMD$. Here, the writing angles remain constant in the reference frame of the optical bench, meaning that the directions of propagation of the reference and image beams are unchanged. Because of this, this method allows for the simplest reading mode, in which the read beam is unchanged, and the HMD is rotated, all while maintaining a constant output direction for the reconstructed image beam. The two principal benefits of this mode are the simplicity in the reading process, whereby the HMD directly controls the information to be read out, and the distinctiveness of the generated gratings, which allow for more optimum utilization of the photosensitive dye.

The second mode leaves the HMD's angle unchanged, instead changing only the rotating mirror's angle. Here, when the mirror is rotated clockwise by $\Delta\angle M_1$, the writing angle relative to the hologram's normal is changed by $\Delta\alpha_{ref} = -2\,\Delta\angle M_1$, leaving $\alpha_{im}$ unchanged. In this mode, the image beam's writing angles are unchanged in the reference frames of the optical bench and the HMD. In this way, the readout is performed by leaving the HMD's angle unchanged, and instead changing the input angle of the reading beam, which once again maintains a constant output direction for the reconstructed image beam. Because the only mobile parameter is the directionality of the reading beam, this mode enables the maximum operational speed for an HMD, as this can be accomplished through use of high-speed acousto-optic deflectors in conjunction with a simple 4f system to allow for independent control of the reading angle and position. The primary drawbacks of this mode are the need for a spacious 4f system and the suboptimal utilization of photosensitive dye due to the unchanging trajectory of the image beam relative to the HMD.

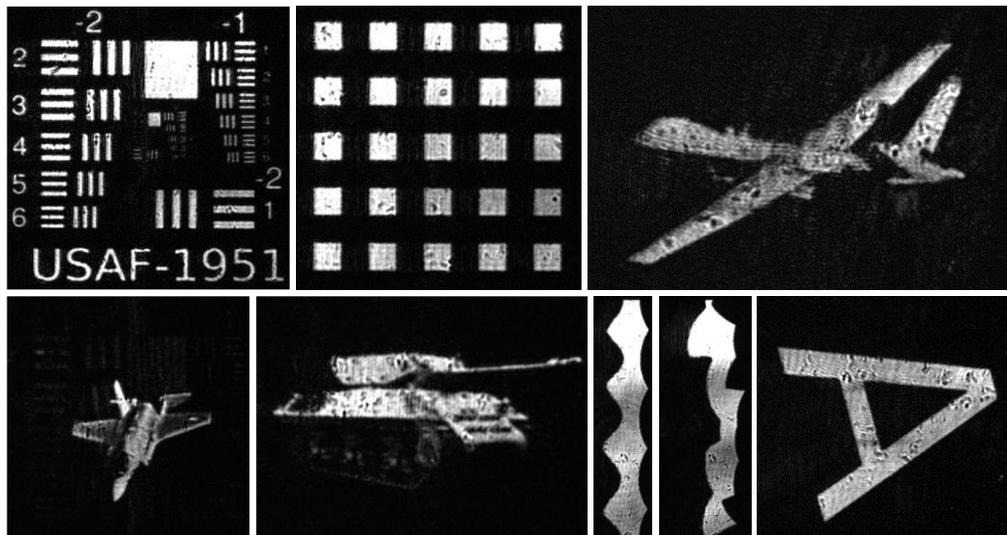

Fig. 2 Typical images retrieved from an HMD with 1,320 total images. Top, left to right: USAF resolution chart, 5x5 grid of squares, MQ-9 Reaper UAV. Bottom, left to right: F-35A fighter jet, Destroyer M10 tank, two artificial PMTs of images with purely real FTs, rotated letter *A*.

Because of the photochemical mechanisms behind grating formation in PQ:PMMA, there is a finite number of gratings that can be written in any one location before the dye has been completely depleted. This can be represented by the M/# of a substrate, which is the ratio of the maximum $\Delta n$ and the value of $\Delta n$ that yields 100% DE in an HOE. Evidently, as holograms are exposed, the dye will be consumed, leaving less and less dye available for subsequent gratings. This would imply that the exposure energy density (Ee) should be increased with subsequent exposures, however this has the added effect of partially erasing previous holograms. Thus, perhaps the most difficult yet important part of writing a large number of stacked holograms in a single location is the selection of an appropriate Ee for each grating. This is often referred to as the writing/exposure schedule. Many strategies exist to define an exposure schedule that will result in a flat DE profile (i.e., all images having the same DE) [19,22], but no single technique will cover all cases. As such, it is important to define the exposure schedule in accordance with the desired reading parameters and application. For this paper, we have selected a slowly increasing exponential writing schedule in order to store 1,320 images across 12 locations (110 images per location) in a 50.4 mm diameter HMD. Fig. 2 shows some representative images retrieved from the HMD and observed on an FPA through a 4-f system.

## 3. SSRI TARGET RECOGNITION THROUGH A HYBRID OPTO-ELECTRONIC CORRELATOR

The HOC was first proposed in [5] and demonstrated in [4]. This device was later augmented to incorporate the PMT for SSRI, as reported in [6] and [9]. The device, a diagram of which is shown in Fig. 3, is composed of three main sections: a phase scanning and stabilization (PSS) arm, a reference image (Ref) arm, and a query image (Que) arm. One of the two image arms, typically the Ref arm, utilizes an HMD as an image source, while the other arm uses an SLM for the same purpose.

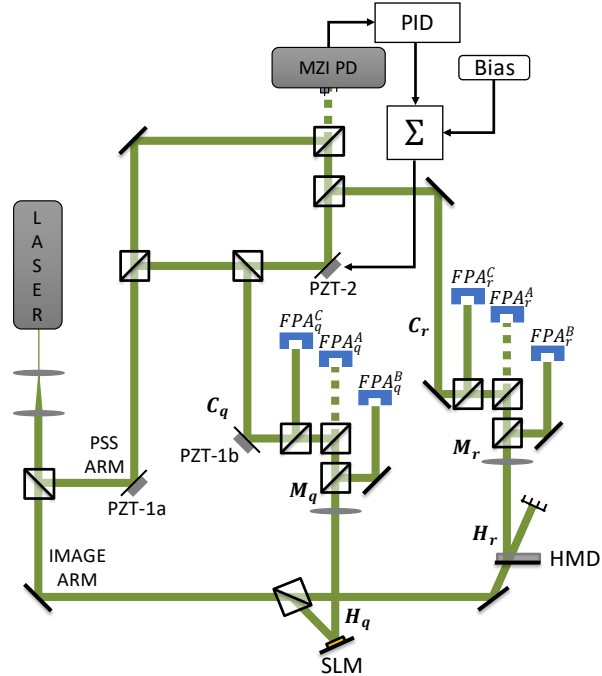

Fig. 3. Schematic diagram of the HOC architecture incorporating an SLM for the query image and an HMD for the reference image.

The image, labeled $H_{r,q}$, where the subscript denotes either the reference or query arm, respectively, is projected onto an expanded 532 nm laser beam and directed towards a biconvex lens placed one focal distance away from the image source. This lens generates the two-dimensional FT of the image, labeled $M_{r,q} = FT\{H_{r,q}\}$, at the opposite focal plane. A beam splitter (BS) placed immediately after the lens splits the beam path so as to generate the FT at two locations: in the reflected path, a mirror redirects the FT towards $FPA_{r,q}^B$; in the transmitted path, another BS is used to combine the FT with an auxiliary plane wave (APW), labeled $C_{r,q}$, which originates from the PSS arm. The interference between the FT and the APW is detected by $FPA_{r,q}^A$. The APW also passes through a third BS which allows it to be detected by $FPA_{r,q}^C$. The system requires the measurement of three FPA intensity signals for each of the two image arms, as indicated by the superscript in the FPA label. These signals are labelled as follows:

$$A_{r,q} = |M_{r,q} + C_{r,q}|^2; B_{r,q} = |M_{r,q}|^2; |C_{r,q}|^2 \qquad (1)$$

where the subscript indicates if the signal was measured from the Ref or Que arm. In this way, $A_{r,q}$ corresponds to the intensity of the complex interference pattern generated between $M_{r,q}$ and $C_{r,q}$, $B_{r,q}$ corresponds to the intensity of the complex FT of the image beam $H_{r,q}$, and $|C_{r,q}|^2$ is the intensity of an APW.

After all measurements are completed, the signals exist in the electronic domain and can be processed either by a computer or by an FPGA. The latter is preferred in order to maximize speed, but the former allows for more refined control during the research stages. The digitized signals are processed in an elementwise manner in order to recover the phase information of the complex wave $M_{r,q}$ from the intensity measurements:

$$S_{r,q} = A_{r,q} - B_{r,q} - |C_{r,q}|^2 = M_{r,q} C_{r,q}^* + M_{r,q}^* C_{r,q} \tag{2}$$

Here, it is clear that the complex information of $M_{r,q}$ is contained within the real variable $S_{r,q}$ alongside the complex information of $C_{r,q}$. The variables $S_r$ and $S_q$ are then multiplied together elementwise:

$$S = S_r \cdot S_q = (M_r C_r^* + M_r^* C_r) \cdot (M_q C_q^* + M_q^* C_q) = \alpha^* M_r M_q + \alpha M_r^* M_q^* + \beta^* M_r M_q^* + \beta M_r^* M_q \tag{3}$$

where we have defined the values $\alpha = C_r C_q$ and $\beta = C_r C_q^*$, which will be constants in the case that $C_r$ and $C_q$ are plane waves. This signal is FT'd to obtain the final output of the HOC. This transformation should ideally be performed optically on an output stage by projecting $S$ on an SLM and passing it through a lens:

$$S_f = FT\{S\} \approx \alpha^* FT\{M_r M_q\} + \alpha FT\{M_r^* M_q^*\} + \beta^* FT\{M_r M_q^*\} + \beta FT\{M_r^* M_q\} \tag{4}$$

Using the well-known relationship between the FT and cross-correlation and convolution operations, we may now write:

$$\begin{aligned} S_f &= \alpha^* T_1 + \alpha T_2 + \beta^* T_3 + \beta T_4 \\ T_1 &= H_r(x,y) \otimes H_q(q,y) \\ T_2 &= H_r(-x,-y) \otimes H_q(-x,-y) \\ T_3 &= H_q(x,y) \odot H_r(x,y) \\ T_4 &= H_r(x,y) \odot H_q(x,y) \end{aligned} \tag{5}$$

However, because the final FT should be carried out in the optical domain, we will only be able to measure its intensity:

$$\begin{aligned} |S_f|^2 &= |\alpha|^2 |T_1|^2 + |\alpha|^2 |T_2|^2 + |\beta|^2 |T_3|^2 + |\beta|^2 |T_4|^2 + (\alpha^2)^* T_1 T_2^* + \alpha^2 T_1^* T_2 + \alpha^* \beta T_1 T_3^* + \alpha \beta^* T_1^* T_3 + \alpha^* \beta^* T_1 T_4^* \\ &+ \alpha \beta T_1^* T_4 + \alpha \beta T_2 T_3^* + \alpha^* \beta^* T_2^* T_3 + \alpha \beta^* T_2 T_4^* + \alpha^* \beta T_2^* T_4 + (\beta^2)^* T_3 T_4^* + \beta^2 T_3^* T_4 \end{aligned} \tag{6}$$

This can be simplified by noting that the convolution and correlation terms must be real numbers, since both $H_r$ and $H_q$ are purely real signals:

$$|S_f|^2 = |\alpha|^2 T_1^2 + |\alpha|^2 T_2^2 + |\beta|^2 T_3^2 + |\beta|^2 T_4^2 + 2T_1 T_2 Re\{\alpha^2\} + 2T_3 T_4 Re\{\beta^2\} + 2Re\{\alpha^* \beta\}(T_1 T_3 + T_2 T_4) + 2Re\{\alpha \beta\}(T_1 T_4 + 2T_2 T_3) \tag{7}$$

One final simplification may be made by noting that some of the terms do not typically overlap in 2D space, and so their elementwise product may be neglected. For images that are *not* symmetric about both *x* and *y*, terms that include $T_1 T_2$, $T_1 T_4$, and $T_2 T_3$ may be taken to be negligible. In addition, the $T_3 T_4$, $T_1 T_3$, and $T_2 T_4$ terms may be discarded when there is a significant spatial shift between the two original images, as this will produce a proportional spatial shift in the $T_1, T_2, T_3$, and $T_4$ terms that will cause them to separate. In the real world, it is extremely rare for an image to be perfectly symmetric about both *x* and *y*, and so with these considerations we may simplify $|S_f|^2$ to be:

$$|S_f|^2_{shifted} \approx |\alpha|^2 T_1^2 + |\alpha|^2 T_2^2 + |\beta|^2 T_3^2 + |\beta|^2 T_4^2 \tag{8}$$

If there is no significant shift, however, the $T_3$ and $T_4$ terms will both be centered about zero, overlapping each other, and as such cannot be discounted:

$$|S_f|^2_{unshifted} \approx |\alpha|^2 T_1^2 + |\alpha|^2 T_2^2 + |\beta|^2 T_3^2 + |\beta|^2 T_4^2 + 2T_3 T_4 Re\{\beta^2\} + 2Re\{\alpha^* \beta\}(T_1 T_3 + T_2 T_4) \tag{9}$$

This shows how it is possible to measure purely real signals and use them to perform a complex-domain calculation. Note, however, the key assumption that was made regarding the APWs: that they are perfectly planar (i.e., that their (x,y) profiles are uniform), and thus can be taken as constants during the final FT. This is not strictly true for real laser beams, as they typically possess a Gaussian profile, yet if the beam is expanded to be much larger than the region that is to be measured and it is clean of defects, then the profile will show little variation and so will have little effect on the FT.

Comparing the two approximations for $|S_f|^2$ shows that there will be a nontrivial dependence on the phases of $\alpha$ and $\beta$ for the case where the original images are unshifted. We can analyze this dependence further by expressing these terms as follows:

$$C_{r,q} = |C_{r,q}|e^{i\Psi_{r,q}}; \quad \alpha = |C_r||C_q|e^{i(\Psi_r+\Psi_q)}; \quad \beta = |C_r||C_q|e^{i(\Psi_r-\Psi_q)} \qquad (10)$$

where $\Psi_{r,q}$ is the phase component of the APWs as measured at their respective FPAs. Both $|C_{r,q}|$ and $\Psi_{r,q}$ are assumed to be spatially constant in the FPA plane. Here, we see that $\alpha$ and $\beta$ have the same magnitude at all times, and so the $T_1^2, T_2^2, T_3^2$, and $T_4^2$ terms are independent of the phases of the APWs. However, the $T_3T_4$ term depends directly on the real part of $\beta^2$, which can be expressed as:

$$Re\{\beta^2\} = |C_r|^2|C_q|^2 \cos(2 \cdot \Delta\Psi) \qquad (11)$$

where $\Delta\Psi = \Psi_r - \Psi_q$. Similarly, the $T_1T_3$ and $T_2T_4$ terms depend directly on the real part of $\alpha^*\beta$, which can be expressed as:

$$Re\{\alpha^*\beta\} = |C_r|^2|C_q|^2 \cos(2 \cdot \Psi_q) \qquad (12)$$

Finally, it is clear that for images that are not symmetric about x and y and do not present a spatial shift relative to each other, the cross-correlation will depend on both the phase difference between the two APWs as well as on the phase of the query APW. As noted earlier, this is only true of images that do not present a significant spatial shift, whereas images that do present such a shift will have a negligibly small $T_3T_4$ term, and so will be unaffected by the phase difference. If we wish to maximize the cross-correlation terms of unshifted images, we may wish to select a value of $2 \cdot \Delta\Psi \approx 2\pi m$, where $m$ is an integer. Alternatively, because we cannot predict whether the input images will have a spatial shift or not, it may be preferable to ensure that the $T_3T_4$ term is small in all cases, so as that its impact in unshifted images can also be considered negligible. In this case, we can select $2 \cdot \Delta\Psi \approx \pi\left(m - \frac{1}{2}\right)$ to minimize the $T_3T_4$ term. The impact of the $T_1T_3$ and $T_2T_4$ terms is non-trivial, as it will depend on both the correlation and convolution terms, and so it may also be preferable to minimize it by setting $2 \cdot \Psi_q \approx \pi\left(m - \frac{1}{2}\right)$.

Stabilizing an optical phase is complicated and unreliable, but many techniques exist to maintain stability for short periods of time. One such method is implemented in the PSS arm of the HOC, which forms a rudimentary discrete optical phase locked loop. Here, a Mach Zehnder interferometer (MZI) is constructed with one mirror mounted on a piezoelectric transducer (PZT) that can precisely change the path length of one arm. A BS placed on a section that is independent of the PZT provides the reference APW, while another BS placed after the PZT provides the query APW. The PZT directly controls the phase of the beam that has reflected off of it, which includes both the section that recombines to form an interference pattern in the MZI, as well as the query APW. A pair of matched detectors are placed after the recombining BS of the MZI in order to measure the intensity at two nearby points in the interference pattern, which will depend directly on the phase difference between the two MZI arms, and so can be controlled through the PZT. Thus, both the interference at the matched detectors, and the phase of the query APW ($\Psi_q$) depend equally on the PZT. A PID loop connected between the matched detectors and the PZT provide a way to set $\Psi_q$ to a random stable value, which can be changed by adjusting the setpoint of the PID. In order to identify the optimum position for $\Psi_q$, which also sets $\Delta\Psi$, a reference correlation with a known result is performed at various PID setpoints. Once the setpoint that yields the expected result is found, additional cross-correlations can be performed.

An alternative and likely superior solution is to force a large spatial shift between both of the input images. This can easily be accomplished by physically positioning the SLM or HMD such that the image is projected far from the center of the input plane. The shift should be large enough that it is impossible for an image from the SLM and an image from the HMD to overlap in this plane. Fig. 4. Shows simulation results of shifted and unshifted correlations alongside their dependence on $\Delta\Psi$. These simulations illustrate the sensitivity of unshifted correlations to the phase of the APW, while simultaneously highlighting the insensitivity of shifted correlations to the same.

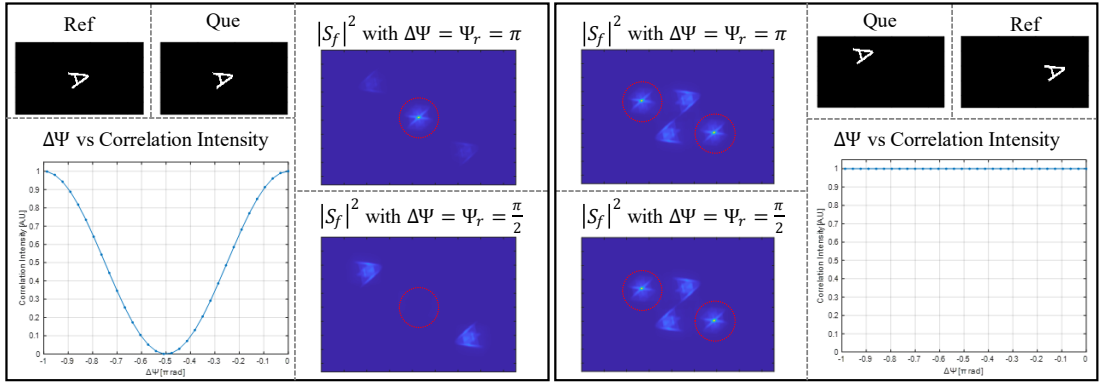

Fig. 4. HOC simulations. Left box: Unshifted images. Right box: Shifted images. The red circles highlight the region where the $T_3^2$ and $T_4^2$ correlation terms are present.

### 3.1  SSRI USING THE POLAR MELLIN TRANSFORM

In the previous section, we showed how the HOC utilizes FTs to calculate the cross-correlation and convolution of two images. Because the FT is inherently shift invariant, the HOC will also have this property. However, the FT is neither scale nor rotation invariant, limiting the feasibility of this device as a target recognition system. This can be overcome by replacing the raw input images with signatures that are independent of scale and rotation. The PMT is an ideal candidate for generating these signatures, as it is closely related to the FT, and so can be partially calculated optically at high speed [6,7]. The PMT can be calculated using the FT through the following process:
1) Calculate the FT of the original image, $G$.
2) Obtain the magnitude of the FT.
3) Define (x=0, y=0) for the FT plane to be located at the center of the FT.
4) Define a grid of pixels with the desired resolution.
5) Calculate the PMT as follows:
   a. The horizontal axis will represent $\rho$, the logarithm of the radius ($r$) of each pixel in the FT:
   $$\rho = \ln\left(\frac{r}{r_0}\right) = \ln\left(\frac{1}{r_0}\sqrt{x^2 + y^2}\right).$$
   where $r_0$ is the minimum radius to be transformed. The horizontal axis will range from $\rho = 0$ ($r = r_0$) to that obtained at the maximum values of x and y ($r = r_{max}$). The calculation is only performed for $r \geq r_0$ in order to avoid divergence of the logarithm operation; this is commonly referred to as circular DC blocking.
   b. The vertical axis will represent $\theta$, the angle of each pixel in the FT: $\theta = \arctan(y/x)$, ranging from 0 to $2\pi$.

The PMT inherits shift invariance from the FT, as any initial shift in $G$ will only result in a phase shift of the FT, which is eliminated after taking its magnitude. Conversely, any rotation in $G$ will result in a proportional rotation of the FT which will convert into a linear shift of the vertical axis $\theta$ of the PMT. Similarly, any scaling in $G$ will result in a radial scaling of the FT, which becomes a linear shift in the horizontal axis of the PMT due to the scaling properties of logarithms: $\rho(a \cdot r) = \ln\left(a\frac{r}{r_0}\right) = \ln(a) + \ln\left(\frac{r}{r_0}\right) = \ln(a) + \rho(r)$, where '$a$' is the scaling factor. Thus, scale and rotation in $G$ are converted into orthogonal linear shifts through the PMT. Finally, if the PMT is used as an input to the HOC such that $H_{r,q} = PMT\{G_{r,q}\}$, the linear shifts that follow from rotation and scaling will simply result in proportionally displaced correlation peaks. This displacement will contain information regarding the scaling and rotation of the original untransformed image, $G$, which can be analyzed as needed. The following section describes a method to perform the PMT in real-time.

## 4. OPTO-ELECTRONIC REAL-TIME PMT GENERATION

The maximum operating speed of the HOC has been estimated to be on the order of a few microseconds [4,5,9], but the actual processing speed reached by recent experiments has been limited due to the lack of specialized components. Furthermore, the use of a computer for pre- and post-processing of the signals results in an additional and significant time delay. One of the most important steps that requires full automation is the pre-processing needed to generate the PMT. To this end, we have designed an opto-electronic PMT pre-processor (OPP) device that combines a field programmable gate array (FPGA), an SLM, and an FPA to perform the PMT. Fig. 5 shows a schematic diagram of the use of this device. In the future, the OPP may be packaged as a single device, but for the moment it has been designed using discrete components on an optical bench.

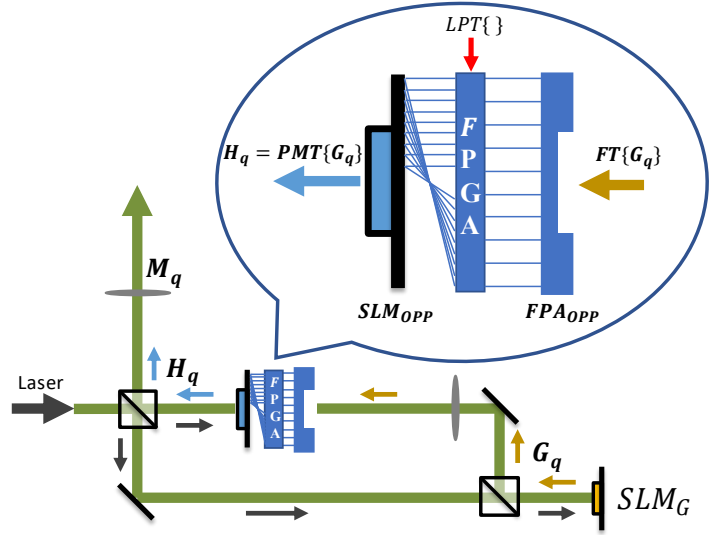

Fig. 5. OPP (Opto-electronic PMT Pre-processor) device diagram. Here, the device is shown as a substitute for the SLM of the query arm in Fig. 3. The dark gray arrows show the path of a simple expanded laser beam. The dark yellow arrows show the path of the original image, $G$, which is projected on $SLM_G$, and its FT, which is detected by the FPA of the OPP. An FPGA performs the LPT on the intensity of the FT and projects the result on another SLM, which serves as the query input to the HOC. The blue arrows denote the path of the input image of the HOC, $H = PMT\{G\}$. The SLMs are taken to be of the reflective type.

In order to perform the PMT, the original image, $G$, is projected on an SLM ($SLM_G$ in Fig. 5) that modulates the intensity of an expanded laser beam. The SLM image is directed through a convex lens that is placed one focal distance away, producing the 2D FT one focal distance away on the opposite side of the lens, where the FPA of the OPP is located. The FPA thus detects $|FT\{G\}|^2$, which is passed on to an FPGA, where the LPT is carried out. Currently, this communication is being done serially, but in the future an application-specific integrated circuit (ASIC) could be constructed to perform this in parallel, drastically increasing performance. Finally, the opto-electronically transformed image $H = PMT\{G\}$ is displayed on the output SLM of the OPP ($SLM_{OPP}$ in Fig. 5).

The proper PMT is the LPT of the magnitude of the FT, but here we use the intensity of the FT instead. This allows us to greatly simplify the OPP, as it reduces the FT-to-PMT conversion to a simple coordinate transformation. Furthermore, the simplification has no impact on the HOC, as the PMT is being used to produce a unique signature that has a one-to-one correspondence with the original image, which is still the case if the intensity of the FT is used. With this simplification in mind, the FPGA must be programmed to perform the LPT in real time. It is important to note that the LPT will always result in the same pixel-mapping for a selected radius $r_0$, independent of the original image. In effect, the output coordinate $(\rho, \theta)$ of a given input coordinate $(x, y)$ is independent of the pixel value at $(x, y)$, and so it can be hard-wired in the FPGA. Fig. 6 shows a simplified example of the coordinate transformation process with and without circular DC blocking.

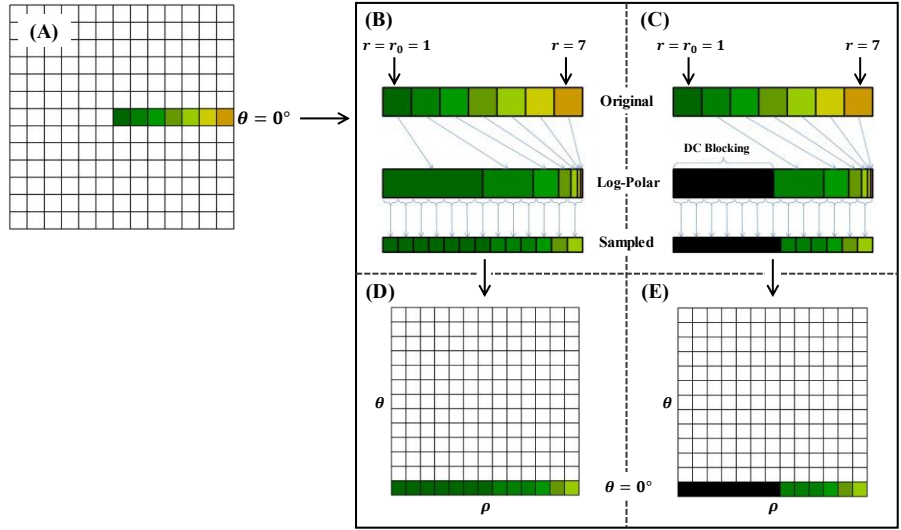

Fig. 6. Simplified example of the LPT process. (A) The coordinate grid of the input $|FT\{G\}|^2$, where the row corresponding to $\theta = 0$ is colored in. (B) The logarithm of the radius is used to obtain scale invariance. The output of the logarithm must be sampled to obtain a consistent pixel count. (C) Same as (B) but with circular DC blocking at $r_0 = 1$, where black signifies that the pixels are left blank. (D)(E) the position of the $(\rho, \theta = 0)$ coordinates in the output grid.

We can take advantage of the lack of dependence on the input image to construct a fast PMT module, whereby the coordinate mapping relationship is pre-calculated on a computer (e.g., in MatLab) and written to the FPGA such that every output pixel is physically connected to its corresponding input pixel, with only a bit-buffer in between to store the greyscale value. This approach yields the fastest possible performance due to the pixel paths operating in parallel, but is impractical on an FPGA for images larger than a few pixels because of the limited number of accessible pins. In the future, an ASIC could be constructed with this architecture to ensure maximum operational speed.

An approach which uses commercially available devices is to store the input data corresponding to $|FT\{G\}|^2$ in fast memory and direct each pixel to its corresponding output coordinate using a pre-calculated mapping written ahead of time on the FPGA. In what follows we will assume the use of random-access memory (RAM) and read-only memory (ROM) modules that allow for the use of two-dimensional memory addresses for the image pixels. In this approach, we use one RAM module to store the original input image, and one ROM module to store the transform coordinate relationship. That is to say, the ROM will store the input coordinates $(x, y)$ at each memory location given by $\rho(x, y)$ and $\theta(x, y)$, where these are integer indices:

$$ROM(\rho_{out}, \theta_{out}) = (x_{in}, y_{in})$$

With this approach, when the FPGA receives a pixel from the FPA with coordinates $(x_{in}, y_{in})$, it can be quickly saved directly to its own address in *RAM* at location $(x_{RAM} = x_{in}, y_{RAM} = y_{in})$:

$$RAM(x_{in}, y_{in}) = |FT\{G\}|^2_{x_{in}, y_{in}}$$

When the FPGA needs to output a pixel with coordinates $(\rho_{out}, \theta_{out})$, it will read the *ROM* at these same coordinates $(x_{ROM} = \rho_{out}, y_{ROM} = \theta_{out})$ to find the *RAM* address with the corresponding input value, at which point that RAM coordinate can be read in order to find the required output value of the pixel. This can be summarized in one abstract equation:

$$RAM\{ROM(\rho_{out}, \theta_{out})\} = |FT\{G\}|^2_{x_{in}, y_{in}}$$

In this scheme, the entirety of the image must be received by the FPGA and stored in RAM. Once reception is complete, the ROM is scanned at the output coordinates, and the RAM is read accordingly to produce the value for the queried output pixel.

To test this architecture, we first use a computer to send and receive the input ($|FT\{G\}|^2$) and output ($PMT\{G\}$) images, respectively, through serial communication. Universal asynchronous receiver/transmitter (UART) was chosen as the communication protocol as it is a simple standard for serialized data that is compatible with USB and is employed in many electronic devices. In the next stage, we will replace the computer with a serialized FPA signal for the input, and an SLM for the output, and in a final stage these devices would be combined using high-speed interconnects. For our initial experiments, the UART baud rate was set to 9,600 bauds/s, which roughly corresponds to 9,600 bits-per-second (bps). The UART standard is limited to ~5 Mbps, which would require ~1.6s for an image with one megapixel of resolution (at 8-bit bit-depth). FPGA development boards are available with other communication standards built in, such as IEEE 802.3, otherwise known as Gigabit Ethernet, which would reduce the transmission time of a one-megapixel resolution image down to ~8ms. It is obvious that in order to reach the maximum theoretical operating speed of the HOC, the FPA and SLM would have to be built into the same ASIC to avoid serial communication altogether.

The FPGA program was developed in VHDL and is modularized into 4 separate parts. These four modules are: UART receive (Rx) module to receive the image from the computer, the RAM module to store the input image, the transform ROM module to store the LPT coordinate relationship, and the UART transmit (Tx) module to transmit the image back to the computer. The block diagram of PMT performing on the FPGA is shown in Fig. 7.

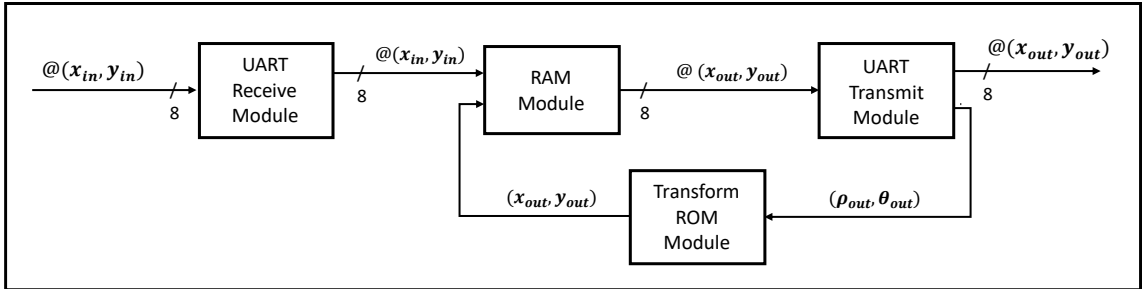

Fig. 7. Modular diagram of the LPT program on an FPGA. The coordinate mapping is pre-calculated on a computer (e.g., in MatLab) and loaded onto the ROM module. Here, the @ symbol preceding a set of coordinates represents the value stored at the corresponding memory address. Coordinates without the @ symbol are memory addresses. In this scheme, the entirety of the $|FT\{G\}|^2$ data must be received and stored in the RAM module prior to transmitting the result.

The Rx module is used to control the process of receiving and decoding the UART signal which has the 8-bit greyscale data of $|FT\{G\}|^2$ for each pixel at coordinates $(x_{in}, y_{in})$. The pixel information is stored directly in the RAM module at the location $(x_{RAM} = x_{in}, y_{RAM} = y_{in})$. The transform ROM module is used to store the coordinate mapping relationship and can only be used for reading. Finally, the Tx module is used to transmit the output image pixel serially, and to control which pixel is being sent at any given time. To do this, the Tx module scans every output pixel coordinate one at a time, consulting the ROM module at the desired $(\rho_{out}, \theta_{out})$ to find the corresponding RAM coordinate, which is read out and finally transmitted as needed.

A computer was used to test the LPT capabilities of the FPGA by sending an artificial image with 120x120 pixels that takes the place of $|FT\{G\}|^2$. Fig. 8 shows the transmitted and received images, which agree with the expected behavior of the LPT.

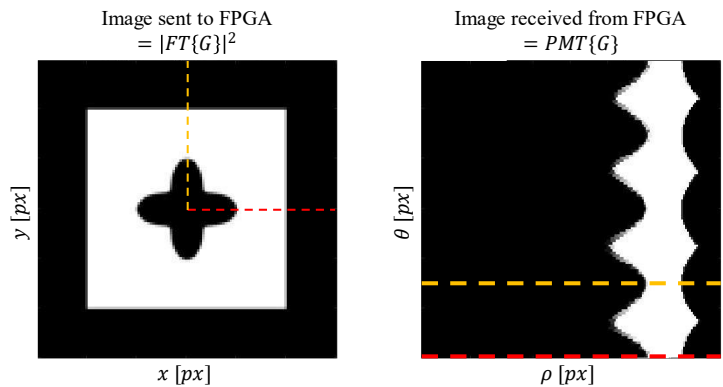

Fig. 8. Left: Image that was transmitted to the FPGA, taking the place of $|FT\{G\}|^2$. Right: Output of the FPGA, which corresponds to the LPT of the input image and so to $PMT\{G\}$. The red dotted line is placed at $\theta = 0°$. The yellow dotted line is placed at $\theta = 90°$.

Further tests remain to be performed with larger images, and the full OPP is yet to be constructed. Yet these results show that the proposed FPGA-based OPP will be able to perform the PMT in a satisfactory manner and at a high speed, opening the doors for high-speed operation of the SSRI HOC.

## 5. EXPERIMENTAL RESULTS

A ~2mm thick HMD was prepared and written with 1,320 images, as described in section 2. The images included pictures, shapes, their respective PMTs, and artificial PMTs that were constructed utilizing arbitrary FT spectra composed of purely real values. These artificial PMTs are ideal for testing the capabilities of both an HMD and the HOC, as they are easy to recognize by eye and have well defined features. Fig. 9 shows the digital versions of four of the images used for the correlation experiments reported here.

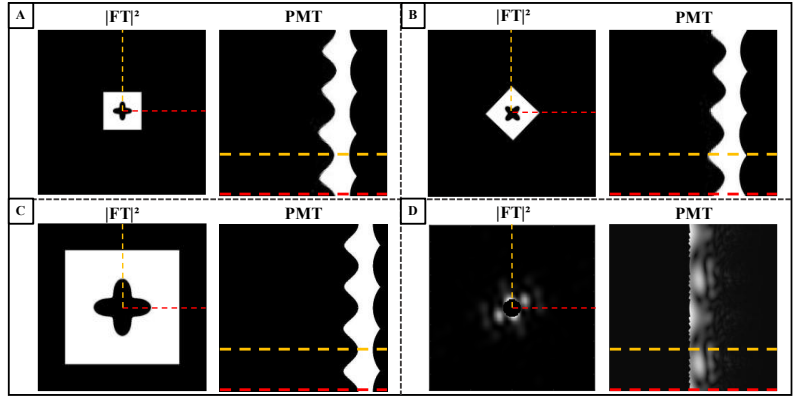

Fig. 9. Digital representation of four of the images used for the experiments reported here. The FTs used as a source for the PMT of sections (A), (B) and (C) were artificially constructed and do not correspond to a real *original* image. (A): Artificial FT composed of a square with a star-shaped hole, along with its PMT. (B): Artificial FT with a rotation of 45° relative to that of (A), along with its PMT. (C): Artificial FT scaled to be 3x larger than that of (A), along with its PMT. (D): Intensity of the FT of a real image of a Soyuz capsule (not shown), along with its PMT. The red dotted lines correspond to $\theta = 0°$. The yellow dotted lines correspond to $\theta = 90°$.

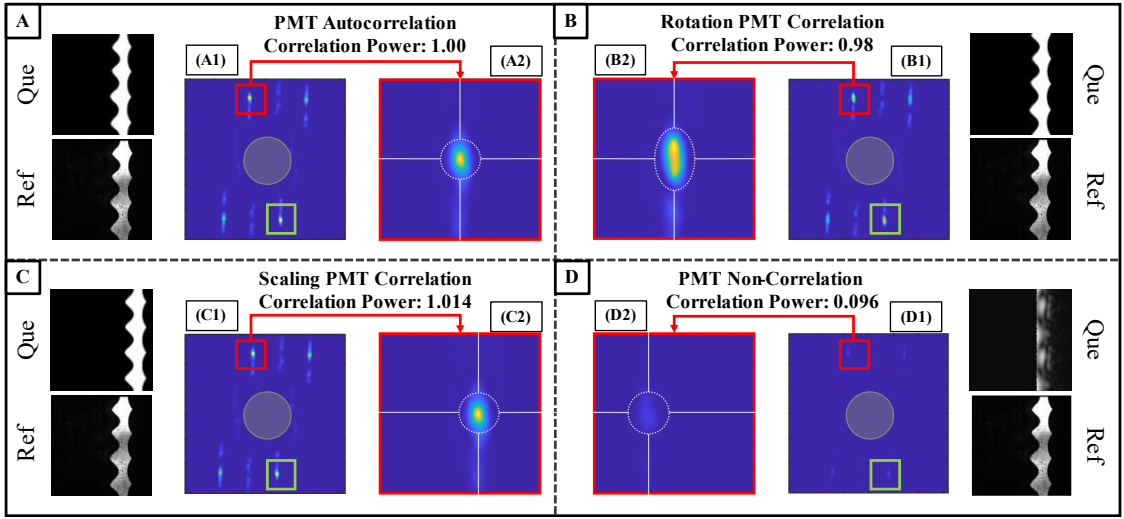

Fig. 10. HOC experiments using PMTs. (A): Autocorrelation. (B): Cross-correlation where the Que PMT is that of Fig. 9(B), representing a rotation of 45°. (C): Cross-correlation where the Que is that of Fig. 9(C), representing a scaling of 1/3. (D): Cross-correlation where the Que PMT is that of Fig. 9(D) and represents an unmatched correlation. (A1,B1,C1,D1): Output of the HOC. The DC region within the dark gray circle has been blocked to allow for the visualization of the correlation and convolution signals. The square regions in red and green correspond to the $T_3^2$ and $T_4^2$ terms, respectively. The squares have been positioned at the same coordinates in all the plots, which corresponds to the peak of the $T_3^2$ term in the autocorrelation result of (A1). (A2,B2,C2,D2): Closeup of the red square regions in (A1),(B1),(C1),(D1), respectively. The white crosshairs emphasize the location of the peak in this region.

The HMD was installed in the reference image arm of the HOC as shown in Fig. 3, where the images were read utilizing the second readout mode described in section 2.1. The SLM was installed in the query image arm and displaced in the input plane by an arbitrarily large amount relative to the HMD in order to produce shifted correlation peaks that are independent of the phases of the auxiliary plane waves (APWs), as described in section 3. Fig. 10 shows the results of four consecutive high-speed experiments whereby the HMD's Ref image was set to be the PMT shown in Fig. 9(A), and the SLM's Que image was varied among the four PMTs shown in Fig. 9. In these experiments, the final FT of $S$ was performed on a computer. Also, due to imperfections in the APWs, it is typical to observe strong noise around the DC region of the final output, which can often be much larger than the correlation signals. For this reason, it is important to block a circular region in the output signal.

In these experiments, the correlation power was defined as the total power in the $T_3^2$ terms of the output signals, which are highlighted in red squares in Figs. 10(A2)(B2)(C2), and (D2) and shown enlarged in Figs. 10(A1)(B1)(C1), and (D1), respectively. Furthermore, the correlation power was normalized to that of the autocorrelation in Fig. 10(A). Figs. 10(B) and (C) yielded powers of 0.98 and 1.014, respectively, demonstrating successful detection of a match despite rotation and scaling. In contrast, Fig. 10(C) yielded a power of 0.096, correctly indicating that there is no match between the two images. In addition, Fig. 10(B2) shows a spreading in the vertical axis of the correlation signal as compared to the autocorrelation of Fig. 10(A2), correctly indicating a shift in the $\theta$-axis of the PMT which is proportional to the rotation of the original image. Likewise, the peak of Fig. 10(C2) presents a displacement in the horizontal axis, correctly indicating a shift in the $\rho$-axis of the PMT corresponding to a scaling factor in the original image.

## 6. CONCLUSIONS AND OUTLOOK

We have demonstrated the use of a high-capacity, high-speed PQ:PMMA based holographic memory device (HMD) with stored polar Mellin transfiorms (PMTs) for shift, scale and rotation invariant (SSRI) correlation using a hybrid opto-electronic correlator (HOC), whereby the inputs are spatially shifted to eliminate dependence on the phase difference between the two auxiliary plane wave (APWs). The HMD contains 1,320 angle multiplexed high quality untransformed images and pre-processed PMTs that can be used for high-speed correlation. The final output of the HOC shows a ~10x higher power at the correlation peak for matches, regardless of shift, scale, or rotation. Furthermore, the position shift of the correlation peaks, which is directly proportional to any rotation and shift in the original images, can clearly be seen and measured, allowing for the approximation of these parameters. In carrying these experiments, we employed a novel approach whereby the HOC inputs are spatially shifted to produce correlation signals without requiring stabilization of optical phases, yielding results that are in good agreement with the theory. Use of this approach vastly simplifies the design and operation of the HOC, while improving its stability significantly. In addition, we showed the functionality of an FPGA-based OPP using commercially available devices. Further work is required in order to implement fully the OPP without suffering from delays due to serialized communication, but this proof of principle is a large step towards real-time generation of PMT signatures.

## 7. FUNDING

The work reported here was supported by AFOSR grant No. FA9550-18-01-0359.